\documentclass{epl}
\graphicspath{{./img/}}
\usepackage{bm}
\usepackage{amsmath,amssymb}

\newcommand{\upi}{\mathrm{i}}
\newcommand{\upe}{\mathrm{e}}

\title{Quantum state preparation in circuit \acro{QED} via Landau-Zener tunneling}
\shorttitle{Quantum state preparation in circuit \acro{QED} via \acro{LZ} tunneling}

\author{
  Keiji Saito\inst{1} \and
  Martijn Wubs\inst{2} \and
  Sigmund Kohler\inst{2} \and
  Peter H\"{a}nggi\inst{2} \and
  Yosuke Kayanuma\inst{3} 
}
\shortauthor{K. Saito \etal}

\institute{
  \inst{1} Department of Physics, Graduate School of Science,
           University of Tokyo - Tokyo 113-0033, Japan \\
  \inst{2} Institut f\"{u}r Physik, Universit\"{a}t Augsburg - D-86135 Augsburg, Germany \\
  \inst{3} Department of Mathematical Science, Graduate School of
           Engineering, Osaka Prefecture University - Sakai 599-8531, Japan
}


\begin{document}
\maketitle

\begin{abstract}
We study a qubit undergoing Landau-Zener transitions enabled by the
coupling to a circuit-QED mode.  Summing an infinite-order
perturbation series, we determine the exact nonadiabatic transition
probability for the qubit, being independent of the frequency of the
QED mode.  Possible applications are single-photon generation and the controllable creation of qubit-oscillator
entanglement.
\end{abstract}


Superconducting loops are promising candidates for solid state qubit
implementations \cite{Nakamura1999a, Wallraff2004a, Blais2004a,
Chiorescu2004a}.  Since the direct observation of
Rabi-oscillations in these systems \cite{Nakamura1999a}, they form the
basis of many experiments on coherent quantum dynamics.  Particularly
interesting are the experiments in circuit quantum electrodynamics (QED)
\cite{Wallraff2004a, Blais2004a, Chiorescu2004a}, which is the solid-state
analogue of a two-level atom in an optical cavity.
Superconducting circuits possess the advantage that many of
their parameters are tunable over a broad range.
This can be exploited for controlling efficiently the qubits.

One particular way of controlling a qubit works by switching the
difference of its diabatic energies from a large negative to a large
positive value, yielding an avoided crossing for the adiabatic
energies.  For sufficiently slow switching, the qubit will
adiabatically follow its instantaneous eigenstates.  In the opposite
limit of fast switching, however, the qubit will abandon the adiabatic
eigenstate and undergo a so-called Landau-Zener (LZ) transition.  The
LZ transition probability can be raised upon increasing the switching
rate \cite{Landau1932a, Zener1932a, Stueckelberg1932a}. 
LZ transitions can be used to effectively control qubit gate operations  \cite{Saito2004a, Hicke2006a} and
to read out qubits \cite{Ankerhold2003a}.  Recently, LZ transitions
have been observed in various experiments with superconducting qubits
\cite{Izmalkov2004a, Ithier2005a, Oliver2005a, Sillanpaa2006b} and
nanomagnets \cite{Wernsdorfer2005a}.

Landau-Zener transitions can also occur for a qubit that is coupled
to a circuit oscillator.  Then, the adiabatic following to the final
ground state takes place even in the absence of a direct coupling
between the two qubit levels, induced instead by the indirect coupling
to the oscillator.  An oscillator in a highly excited coherent state
can be described classically, yielding a non-monotonic behavior of the
LZ transition probability versus the coupling strength
\cite{Wubs2005a}.  

Experimentally relevant is also the opposite limit
in which the oscillator is initially in its ground state.  
Here, we shall investigate the Landau-Zener dynamics of such a
qubit-oscillator setup and make specific suggestions for exploiting
it to manipulate the quantum states of the qubit and the oscillator.
We derive an exact analytical expression for the transition
probabilities and we propose the creation of single circuit photons.

%
In tunneling representation, a circuit QED setup is  described by
the Hamiltonian \cite{Blais2004a, Wallraff2004a}
\begin{equation}
\label{ham}
H(t) = -\frac{ E_\mathrm{el} }{ 2 } \sigma_{x}
    -\frac{E_\mathrm{J}(t)}{2}\sigma_{z}
    + \hbar \Omega b^{\dagger} b
    + \gamma  ( b^{\dagger} + b ) \left[\sigma_{x} +1- 2 N_\mathrm{g}\right]
\end{equation}
for the qubit, the circuit oscillator, and their mutual coupling.
The cavity is modelled as a harmonic oscillator with the
annihilation and creation operators $b$ and $b^\dagger$, while
$\sigma_{x}$ and $\sigma_z$ denote Pauli matrices.  The electrostatic
energy $E_\mathrm{el} = 4E_\mathrm{c}[1-2 N_\mathrm{g}]$ is determined
by the charging energy $E_\mathrm{c}$ and the tunable gate charge
$N_\mathrm{g}$.  The tunable flux $\Phi(t)$ penetrating the superconducting loop
will be used to drive the qubit. The flux controls the Josephson energy $E_\mathrm{J}(t) =
E_\mathrm{J,max}\cos[\pi\Phi(t)/\Phi_0]$, where $\Phi_{0}$ is the flux
quantum.  The two-level approximation underlying the Hamiltonian
\eqref{ham} is valid in the charge regime $E_\mathrm{c}\gg  E_\mathrm{J}$.
In order to minimize decoherence, one typically operates the qubit at
the optimal working point $N_\mathrm{g} = \frac{1}{2}$, so that
$E_\mathrm{el}=0$ \cite{Vion2002a}.
Henceforth, we restrict ourselves to this optimal working point.  The
LZ dynamics can then be realised by switching the flux $\Phi(t)$ in such a way that
$E_\mathrm{J}(t) = -vt$, with $v>0$. The duration of this linear sweep
has to be long enough, so that transition probabilities have converged
and the finite time interval can be extended to
$t=-\infty\ldots\infty$ in calculations describing the dynamics.
Since the energy splitting of the qubit can even vanish, it is
essential to treat the qubit-oscillator coupling beyond rotating-wave
approximation.
The temperatures in circuit QED experiments \cite{Wallraff2004a}
justify the assumption that both the qubit and the oscillator are
initially in their ground states, i.e. $|\Psi(-\infty)\rangle =
|{\uparrow},0\rangle$, where $\sigma_z |{\uparrow}\rangle =
|{\uparrow}\rangle$.

\section{Nonadiabatic transitions}
For $E_\mathrm{el}=0$, the states $|{\uparrow}\rangle$ and
$|{\downarrow}\rangle$ are eigenstates of the qubit Hamiltonian
$\frac{1}{2}vt\sigma_z$
and any transition between these states can only result from the
coupling to the oscillator.  The central quantity of interest is the
probability $P_{\uparrow\to\downarrow}(t) = 1  -P_{\uparrow\to\uparrow}(t) $ that the qubit has flipped to the state
$|{\downarrow}\rangle$.    In the following, we derive an exact
expression for $P_{\uparrow\to\uparrow}(\infty) = \sum_n
|\langle{\uparrow},n| U(\infty,-\infty) |{\uparrow},0\rangle |^2$
where $U(t,t')$ denotes the time-evolution operator and $|n\rangle$ an
oscillator eigenstate.
We start by a transformation to an interaction picture with respect
to the uncoupled qubit and oscillator, $U_0(t) = \exp(-\upi\Omega
b^\dagger b t) \exp(-\frac{\upi}{4\hbar}vt^2 \sigma_z)$.  This yields
the interaction-picture Hamiltonian
\begin{equation}
\label{Hint}
\tilde H(t) = \gamma  (b^\dagger e^{\upi\Omega t} + b\, e^{-\upi\Omega t})
              \exp\Big(-\frac{\upi}{2\hbar}vt^2\sigma_z\Big) \sigma_x .
\end{equation}
A perturbation expansion of the probability
amplitude $A_n = \langle{\uparrow},n| U(\infty,-\infty)
|{\uparrow},0\rangle$ results in the series $A_n = \sum_{k=0}^\infty
(-\upi/\hbar)^{2k} a_{nk}$, with the $2k$-th order contribution $a_{nk}$ equal to
\begin{equation}
\label{ank}
\sum_{\lambda_{2k}\cdots\lambda_1}
\gamma^{2k} C_{nk}(\{\lambda \})
\int_{-\infty}^\infty \upd t_{2k} \int_{-\infty}^{t_{2k}}
\upd t_{2k-1} ... \int_{-\infty}^{t_2} \upd t_1
\exp\Big[\upi \sum_{\ell=1}^{2k}\left( \Omega \lambda_\ell t_\ell
         +\frac{v}{2\hbar} (t_{2\ell}^2 - t_{2\ell-1}^2) \right)\Big].
\end{equation}
Since each $\tilde H(t)$ flips the qubit exactly once, only even orders
of $\gamma$ appear in $A_n$.
The coefficients $\lambda_\ell=\pm 1$, for $\ell=1,\ldots, 2k$, stem from
the sign in the time-dependent phase of the creation and annihilation
operators; $C_{nk}(\{\lambda \})=C_{nk}(\lambda_{2k},\ldots,\lambda_1) = \langle
n|\cdots|0\rangle$, where the dots denote the combination of $2k$
creation and annihilation operators $b$ and $b^\dagger$ that
corresponds to the sequence $\lambda_{2k},\ldots,\lambda_1$.
An important simplification of the $\lambda$-summation results from
the fact that $C_{nk}=0$ whenever more annihilation than
creation operators act on the oscillator ground state $|0\rangle$.
Thus, we need to consider only those $\lambda$-sequences that  fulfill the relation
\begin{equation}
\sum_{\ell'=1}^{\ell} \lambda_{\ell'} \geq 0, \qquad \forall\,\ell \leq 2k.
\label{cond1}
\end{equation}

For the further evaluation, we substitute in Eq.~\eqref{ank} the times
$t_\ell$ by the time differences $\tau_\ell = t_{\ell+1}-t_\ell$, for
$\ell = 1,\ldots,2k-1$, where $t=t_{2k}$.  Thus, we insert $t_\ell = t -
\sum_{\ell'=\ell}^{2k-1} \tau_{\ell'}$, so that the integral in
\eqref{ank} becomes
\begin{equation}
\label{int}
\int_{-\infty}^\infty \upd t \int_0^\infty \upd\tau_{2k-1}\ldots \upd\tau_1
\exp\Big[\upi\Omega\sum_{\ell=1}^{2k} \lambda_\ell
\Big(t-\sum_{\ell'=\ell}^{2k-1} \tau_{\ell'} \Big)
+\frac{\upi v}{2\hbar} \sum_{\ell=1}^k \Big\{ 2\tau_{2\ell-1}
\Big(t - \sum_{\ell'=2\ell}^{2k-1} \tau_{\ell'} \Big) - \tau_{2\ell-1}^2
\Big\}
\Big] .
\end{equation}
The $t$-integration results in the delta function
\begin{equation}\label{deltafunction}
2\pi\, \delta\Big( \frac{v}{\hbar}\sum_{\ell=1}^k \tau_{2\ell-1} + \Omega
\sum_{\ell=1}^{2k} \lambda_\ell\Big) .
\end{equation}
From the inequality \eqref{cond1} it follows that the second sum in the
argument of the delta function is non-negative.  Because the
integration interval of all $\tau_\ell$ is $[0\ldots\infty)$, any
non-zero contribution to the integral \eqref{int} comes from $\tau_1
= \tau_3 = \ldots = \tau_{2k-1} = 0$.  Hence, the
integral over the time differences $\tau_2, \tau_4, \ldots, \tau_{2k-2}$
must yield a distribution proportional to
$\delta(\tau_1)\,\delta(\tau_3)\cdots \delta(\tau_{2k-1})$.
Evaluating the integrals over all $\tau_{2\ell}$ separately,
one finds that such a distribution is obtained only if
$\sum_{\ell'=1}^{2\ell}\lambda_{\ell'} = 0$ for all
$\ell=1,\ldots, k-1$.  These $k-1$ relations together with the delta
function~\eqref{deltafunction} lead to the conditions
$\lambda_{2\ell}+\lambda_{2\ell-1} = 0$ for all
$\ell=1,\ldots,k$.  In combination with Eq.~\eqref{cond1} they imply that only
those integrals with $\lambda_\ell = -(-1)^\ell$ are non-vanishing.
In other words, we obtain the selection rule that to the final
occupation probability only those processes contribute in which the
oscillator jumps (repeatedly) from the state
$|0\rangle$ to the state $|1\rangle$ and back.  Hence the only
relevant $C_{nk}$ reads $C_{nk}(-1,1,\ldots,-1,1) = \langle
n|(bb^\dagger)^k|0\rangle = \delta_{n,0}$.

The remaining multiple integrations are performed as detailed in
Refs.~\cite{Kayanuma1984a, Grifoni1998a} and result in  $a_{nk} = \delta_{n,0}
(\pi\hbar/v)^k/k!$, so that $A_n = \delta_{n,0}
\exp(-\pi\gamma^2/\hbar v)$.  This implies that the
oscillator returns to its ground state provided the qubit ends up in state
$|{\uparrow}\rangle$.  Thereby, we arrive at a first central result, namely
the exact transition probability
\begin{equation}
\label{centralresult}
P_{\uparrow\to\downarrow}(\infty)
= 1-P_{\uparrow\to\uparrow}(\infty)
= 1-\upe^{-2\pi\gamma^2/\hbar v} \;.
\end{equation}
Most surprisingly, it does {\em not} involve the oscillator frequency
$\Omega$. Because of this, Eq.~(\ref{centralresult}) allows a simple
interpretation: if the Josephson energy is switched slowly (and the
condition for that is $\hbar v \ll \gamma^{2}$), then the qubit will
follow the adiabatic ground state which at large times is the state
$|{\downarrow}\rangle$.  For large $\hbar v/\gamma^{2}$, the qubit
will remain in state $|{\uparrow}\rangle$, corresponding to a
nonadiabatic transition. Notice the important difference to the
standard LZ problem that 
an intrinsic coupling between the qubit levels is replaced here by the qubit-oscillator coupling $\gamma$.

Since only the states $|{\uparrow},0\rangle$ and
$|{\downarrow},1\rangle$ contribute to the perturbation series for
$P_{{\uparrow}\to{\uparrow}}(\infty)$, it is tempting to conclude
\textit{a posteriori} that for the qubit-oscillator coupling, a
rotating-wave approximation (RWA) is justified.  This would restrict the
whole dynamics to the two mentioned states and indeed the same
probability $P_{{\uparrow}\to{\uparrow}}(\infty)$ would be obtained.
However, below we will find that at intermediate times, states
$|{\uparrow},2n\rangle$ and $|{\downarrow},2n{+}1\rangle$ with $n>0$
can be considerably populated---the latter even at $t\to\infty$.
This demonstrates the surprising fact that the RWA yields the correct
transition probability \eqref{centralresult} even when the RWA is not justified.

\begin{figure}[t]
\onefigure{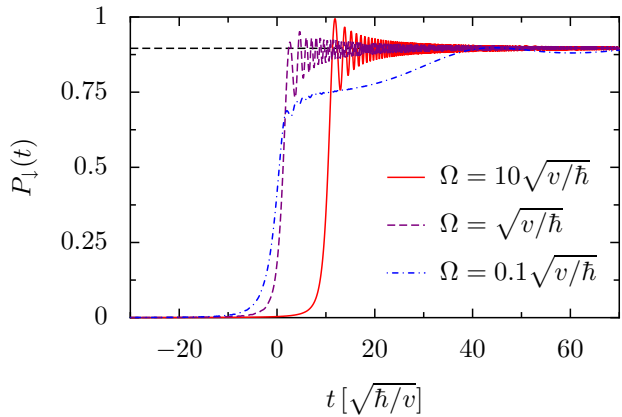}
\caption{
Landau-Zener dynamics for the coupling strength $\gamma=0.6\sqrt{\hbar v}$
for various cavity frequencies $\Omega$.  The dashed line marks the
$\Omega$-{independent}, final probability \eqref{centralresult} to
which all curves converge.}
\label{fig:one-osc}
\end{figure}

%
In order to gain information about
the dynamics at intermediate times, we numerically integrated the
Schr\"odinger equation for the Hamiltonian \eqref{Hint}.
The time evolution of the probability that the qubit is in
state $|{\downarrow}\rangle$ is depicted in Fig.~\ref{fig:one-osc}.  It
demonstrates that at intermediate times, the dynamics depends strongly
on the oscillator frequency $\Omega$, despite the fact that this is not the
case for long times. 
For a large
oscillator frequency, $P_{\uparrow\to\downarrow}(t)$ resembles the
standard LZ transition with a time shift $\hbar\Omega/v$.
 
If the system starts out not in the ground state but in a state
$|{\uparrow},n\rangle$ with $n>0$, some transition probabilities can
still be obtained analytically.  We find $P_{\uparrow n \rightarrow
\uparrow (n+m)}=0$ for all $m>0$, being a special case of the
selection rule derived in Refs.~\cite{Sinitsyn2004a, Volkov2005a}.
Moreover, $P_{\uparrow n \rightarrow \uparrow n}=\exp\left[-2\pi
(2n+1)\gamma^{2}/\hbar v\right]$, indicating that
$|{\uparrow},0\rangle$ is the most stable state.

\section{Manipulation of the oscillator state}
\begin{figure}[t]
\onefigure{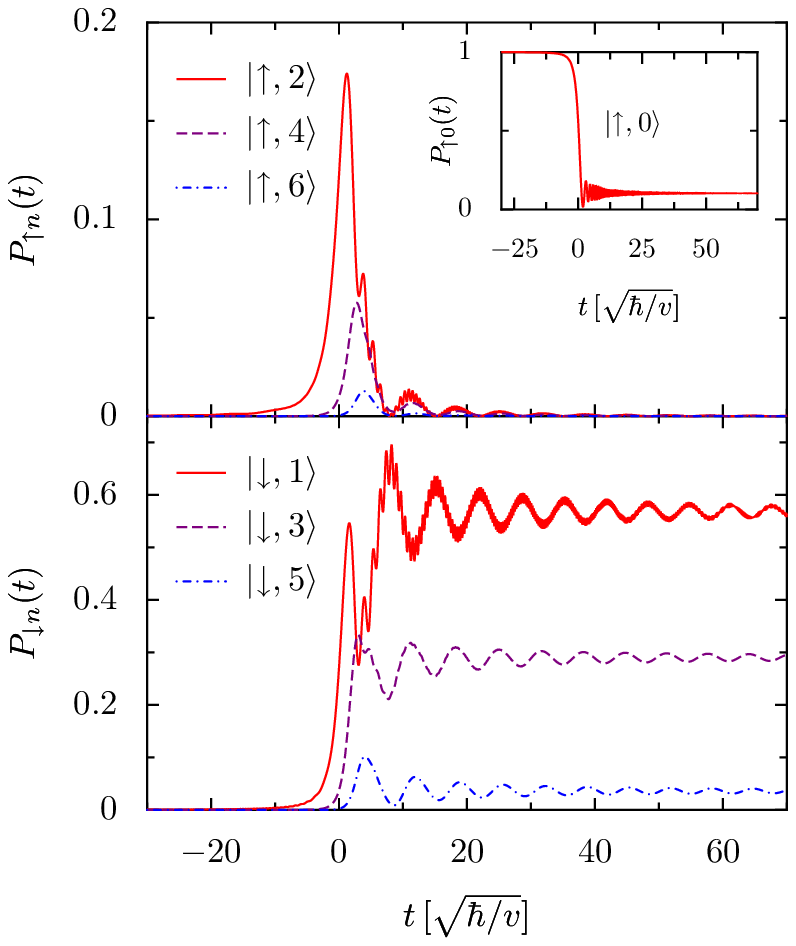}
\caption{Population dynamics of individual
qubit-oscillator states for a coupling strength $\gamma=0.6\sqrt{\hbar
v}$ and oscillator frequency $\Omega=0.5 \sqrt{v/\hbar}$.}
\label{fig:updown}
\end{figure}

Having studied the qubit dynamics, we next focus on
the modification of the oscillator state.
Since we start out in the ground state $|{\uparrow},0\rangle$ and the
Hamiltonian \eqref{Hint} correlates every creation or annihilation of
a photon with a qubit flip, the resulting dynamics is restricted to
the states $|{\uparrow},2n\rangle$ and $|{\downarrow},2n{+}1\rangle$.
Figure~\ref{fig:updown} reveals that the latter states survive for long
times, while of the former states only $|{\uparrow},0\rangle$
stays occupied, as it follows from the relation that $A_n
\propto\delta_{n,0}$, derived above. Thus, the final state exhibits a
peculiar type of entanglement between the qubit and the oscillator, and
can be written as
\begin{equation}
|\Psi(\infty)\rangle
= \sqrt{1-P_{{\uparrow}\to{\downarrow}}(\infty)}\,|{\uparrow},0\rangle
 +\sqrt{P_{{\uparrow}\to{\downarrow}}(\infty)}\,
  \big(c_1|{\downarrow},1\rangle + c_3|{\downarrow},3\rangle
       +\ldots\big) ,
\end{equation}
where $|c_1|^2+|c_3|^2+\ldots = 1$.  This implies that only odd-photon
states are occupied if the qubit ends in $|{\downarrow}\rangle$,
representing a highly nonclassical oscillator state.  Qubit and
oscillator end up fully entangled, in the sense that after tracing out
the oscillator states, no coherence between the qubit states
$|{\uparrow}\rangle$ and $|{\downarrow}\rangle$ is left.

While $P_{{\uparrow}\to{\downarrow}}(\infty)$ is determined by the
ratio $\gamma^2/\hbar v$, the coefficients $c_{2n+1}$ depend also on
the oscillator frequency.
Experiments in circuit QED have been performed for
$\gamma\ll\hbar\Omega$ in the coherent
regime \cite{Wallraff2004a}.  There, $c_1\approx 1$ to a very good
approximation. Hence one can control via $v$ the
final state to be any superposition of $|{\uparrow},0\rangle$ and
$|{\downarrow},1\rangle$.
In particular, in the adiabatic limit $v\hbar/\gamma^2 \ll 1$, the
final state becomes $|{\downarrow},1\rangle$.
This has the important physical implication of
the creation of exactly one photon in the cavity, triggered by
a Landau-Zener transition.  In an experiment, the photon will
subsequently leak out of the cavity.
By exploiting these two processes, we propose the following four-step
LZ cycle for single-photon generation: The first step is single-photon
generation in the cavity via the adiabatic LZ
transition $|{\uparrow},0\rangle \rightarrow |{\downarrow},1\rangle$, brought about by switching the
Josephson energy sufficiently slowly. Second, the photon is released from the cavity via the (controlled) cavity
decay $|{\downarrow},1\rangle \rightarrow |{\downarrow},0\rangle$. In
the  third step, another individual photon is generated via the
reverse LZ sweep $|{\downarrow},0\rangle \rightarrow
|{\uparrow},1\rangle$. Fourth and finally, a further photon decay
completes the cycle.

This scheme for repeated photon generation via Landau-Zener cycles
makes use of two advantageous properties of circuit QED: first, the
artificial atom is fixed at an antinode of the cavity, so that the
atom-cavity coupling remains at a constant and high value. Second, qubits are highly tunable so that LZ sweeps can be
made from minus to plus an ``atomic'' frequency, and back. 
The outlined scheme presents an alternative to the proposals for
single-photon generation put forward in Refs.~\cite{Liu2004a,
Mariantoni2006a}.  The main advantage of the present scheme is its
robustness against parameter variations when operating in the regime
$\gamma\ll\hbar\Omega$.  

\section{Experimental realisation} 
\begin{figure}[t]
  \onefigure{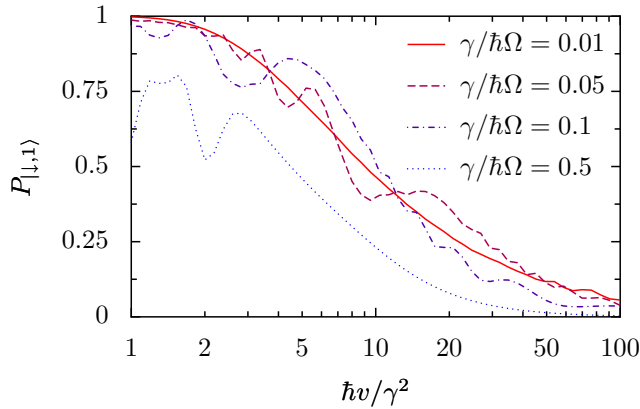}
  \caption{Probability of single-photon generation
  $P_{|{\downarrow},1\rangle}$ as a function of $\hbar v /\gamma^{2}$,
  for  LZ sweeps within the finite time interval $[-T
  ,T]$ with $T>T_{\mathrm{min}}$ chosen such that $v
  T =
  3\hbar \Omega/2$. The initial state is $|{\uparrow,0}\rangle$.
  Shown probabilities are averaged within the time interval
  $\frac{29}{20}\hbar\Omega/v$ and $\frac{3}{2}\hbar\Omega/v$, 
  whereby the small and fast oscillations that are typical for the  tail of a LZ transition are averaged out.}
\label{fig:P_single}
\end{figure}

In practice, the cavity frequency $\Omega$ and the
qubit-oscillator coupling $\gamma$ are determined by the design of the
setup, while the Josephson energy can be switched at a controllable
velocity $v$ --- ideally from $E_\mathrm{J}=-\infty$ to
$E_\mathrm{J}=\infty$.  In reality, however, $E_\mathrm{J}$ is bounded
by $E_\mathrm{J,max}$ which is determined by the critical current. The condition $E_\mathrm{J,max }> \hbar \Omega$ is required so that the qubit comes into resonance with the oscillator sometime during the sweep.
Moreover, inverting the flux through the superconducting loop requires
a finite time $2 T_{\mathrm{min}}$, so that $v$ cannot exceed $v_\mathrm{max} =
E_\mathrm{J,max}/2 T_{\mathrm{min}}$. 
In order to study under which conditions the finite initial and
final times can be replaced by $\pm\infty$, we have numerically integrated the
Schr\"odinger equation in a finite time interval
$[-T,T]$. Results are presented in Fig.~\ref{fig:P_single}.
The first three curves in Fig.~\ref{fig:P_single} correspond to the
experimentally relevant regime where $\gamma \ll \hbar\Omega$. Here
the  essential dynamics
is a LZ transition from $|{\uparrow},0\rangle$ to
$|{\downarrow},1\rangle$  around the time $\hbar\Omega/v$.  Especially for the smallest coupling
constants shown, the probability for single-photon generation
practically equals $P_{{\uparrow \rightarrow \downarrow}}(\infty)$
given by
Eq.~\eqref{centralresult}.  Thus we find that
finite-time effects do not play a role as long as $\gamma
\ll\hbar\Omega$. Our predicted transition
probabilities based on analytical results for infinite
propagation time are therefore useful to describe 
the finite-time LZ sweeps. Figure~\ref{fig:P_single} also illustrates
that the probability for single-photon production is highest in the
adiabatic regime $\hbar v/\gamma^{2}\ll 1$. Here the 
typical duration of a LZ transition is  $2 \gamma /v$
\cite{Vitanov1999a,Wubs2005a}.   So in the regime of interest, the sought condition for a ``practically
infinite time interval'' is $v T = E_\mathrm{J,max} >
\hbar\Omega+2\gamma$.  For the unrealistically large
qubit-oscillator coupling $\gamma/\hbar\Omega = 0.5$,
reliable single-photon generation is less probable. This is so because (i) the LZ transition is incomplete
within $[-T,T]$; (ii) more than two
oscillator levels take part in the dynamics and more than one photon
can be generated, as depicted in Fig.~\ref{fig:updown}; and (iii) the
approximation of the instantaneous ground state at $t=-T$
by $|{\uparrow,0}\rangle$ is less accurate.

For the setup of Refs.~\cite{Wallraff2004a, Blais2004a}, a typical
cavity frequency is $\Omega = 2\pi\times 10^9\mathrm{Hz}$.  
The sign of the initial Josephson energy $E_\mathrm{J,max} = 2\pi\hbar\times
10^{10}\mathrm{Hz}$ can be inverted within $T = 1\mu\mathrm{s}$
so that $v_\mathrm{max} = 2\pi \hbar\times 10^{16}\mathrm{s}^{-2}$. For the qubit-oscillator coupling strength we assume $\gamma/2 \pi \hbar = 3\times 10^6\mathrm{Hz}$. Notice that $\gamma \ll \hbar \Omega$, so that generation of more than one photon is negligible. Then, by choosing $v = 0.05 v_{\mathrm{max}}$, one finds  $|c_{1}|^{2}> 1-10^{-7}$ and a superposition with equal weights for which $P_{{\uparrow}\to{\uparrow}}(\infty)= |c_{1}|^{2}P_{{\uparrow}\to{\downarrow}}(\infty)=0.5$ is obtained. By choosing $v=v_{\mathrm{max}}$ instead, a nonadiabatic transition occurs with the
probability $P_{{\uparrow}\to{\uparrow}}(\infty)\approx 0.97$, while the remaining probability again corresponds to single-photon generation. As another extreme case, for the much slower sweep velocity $v=0.01 v_\mathrm{max}$, one is in
the adiabatic limit in which a single photon is created with a
probability of $|c_{1}|^{2} P_{{\uparrow}\to{\downarrow}}(\infty)= 0.97$.
Thus for the final state, one can obtain  any desired superposition
of the states $|{\uparrow},0\rangle$ and $|{\downarrow},1\rangle$ by
choosing a proper value of $v$.

%
In summary, we have shown that the coupling of a qubit to
a circuit cavity mode induces Landau-Zener transitions upon variation
of the penetrating flux.  For an oscillator initially in its ground state,
we derived an exact closed expression for the transition
probability which possesses an appealing form: it only depends on the
interaction strength between the qubit and the oscillator and, in
particular, it is independent of the actual oscillator frequency $\Omega$.
Thus, the LZ dynamics can be manipulated via the velocity at which the
qubit levels cross.  For a circuit QED setup, we found that both
nonadiabatic and adiabatic sweeps are feasible and can be exploited
for quantum state preparation. Repeated  adiabatically
slow Landau-Zener sweeps allow the controlled and robust creation of single
photons.

\acknowledgements

We thank A. Wallraff for helpful discussions.
This work has been supported by the Freistaat Bayern via the ``Quantum
Information Highway A8'',  the DFG through SFB\,631, and a
Grant-in-Aid for Scientific Research of Priority Area
from the Ministry of Education, Sciences, Sports, Culture and Technology of
Japan (No.\ 14077216).


\end{document}